\documentstyle[12pt]{article}
\begin{document}

\begin{flushright}
MPI-Ph/93-52\\
TUM-T31-45/93\\
July 1993\\
\end{flushright}
%\vspace{70mm}
\begin{center}
\large {\bf A 1993 Look at the Lower Bound on \\
the Top Quark Mass from CP Violation}\\
\mbox{ }\\
%\vfill
\normalsize
{\bf Andrzej J. Buras}\\

\bigskip
Technische Universit\"at M\"unchen, Physik Department\\
D-85748 Garching, Germany

and

Max Planck Institut f\"ur Physik \\
D-80805 M\"unchen, P.O. Box 401212, Germany \\
%%\vspace{20mm}
\vspace{4cm}

{\bf Abstract}\\
\end{center}
We point out that the lower bound on $m_t$ from the
CP violation parameter $\epsilon_K$ has increased considerably.
Using Wolfenstein parametrization of the CKM
matrix we derive an analytic expression for this bound
as a function of $\mid V_{cb}\mid,
\mid V_{ub}/V_{cb}\mid$ and the non-perturbative parameter
$B_K$. For $B_K \leq 0.80, \mid V_{cb}\mid \leq 0.040$
and $\mid V_{ub}/V_{cb}\mid \leq 0.10$
we find $m_t > 130 GeV$. However, for $B_K \leq 0.70,
\mid V_{cb}\mid \leq 0.038 $ and $\mid V_{ub}/V_{cb}\mid
\leq 0.08$ the bound is raised to $m_t > 205~GeV$.
The lower bound on $m_t$ from $B^o-\bar B^o$ mixing
is also reconsidered.
\newpage

Ten years ago Ginsparg, Glashow and Wise \cite{PHG} calculated
a lower
bound on the top quark mass $(m_t)$ by analysing the
CP violation parameter $\epsilon_K$ in the standard model.
Subsequently a more detailed analysis has been done by
S{\l}ominski, Steger and the present author \cite{AJB}.
The lower
bound on $m_t$ considered in \cite{PHG,AJB}
takes the general form
\begin{equation}
m_t \geq F (\mid V_{cb}\mid, \mid V_{ub}/V_{ub}\mid, B_K)
\label{F1}
\end{equation}
where $V_{ij}$ are the elements of the
Cabibbo-Kobayashi-Maskawa
matrix and $B_K$ is a non-perturbative
parameter to be specified below. The function $F$ increases
with decreasing $\mid V_{cb}\mid, \mid V_{ub}/V_{cb}\mid$
and $B_K$.

In 1983 the values of $\mid V_{cb}\mid, \mid
V_{ub}/V_{cb}\mid$ and of $B_K$ have been poorly known
and no useful bound on $m_t$ could be obtained. During
the last ten years the experimental determinations of $\mid
V_{cb}\mid $ and $\mid V_{ub}/V_{cb}\mid$ have been considerably
improved and some progress has been made in calculating
$B_K$. It is of interest then to reanalyse the bound in view
of these developments and in view of the possibility that the
top quark could be discovered soon.

In fact the present ``best'' values for $\mid V_{cb}\mid,
\mid V_{ub}/V_{cb}\mid $ and $B_K$ are substantially lower
than the corresponding values of 1983. Consequently the
lower bound on $m_t$ from $\epsilon_K$ can be considerably
improved.

We note:

\begin{itemize}
\item The new experimental results on exclusive semi-leptonic
$B$-decays \cite{SLS} combined with the increased $B$-meson life-time
measured at LEP and by CDF at Tevatron
$\tau _B = 1.49 \pm 0.03 ps$ \cite{NA} give with the help of the
Heavy Quark Effective Theory \cite{NEU,MN}
\begin{equation}
\kappa \equiv \left| V_{cb} \right|
\left(\frac{\tau_B}{1.5 ps}\right)^{1/2}
= 0.038 \pm
0.006
\label{F2}
\end{equation}
Similar results can be found in \cite{BB,ST}. This value
should be compared with $\mid V_{cb}\mid \approx
 0.05$ of 1983.
\item The most recent determinations of
$\mid V_{ub}/V_{cb} \mid$ give typically \cite{BBD,ART}
\begin{equation}
\left|\frac{V_{ub}}{V_{cb}} \right| = 0.08 \pm 0.02
\label{F3}
\end{equation}
whereas in 1983 values as high as $\mid V_{ub}/V_{cb}\mid
\simeq 0.20$ were possible.
\item The extensive calculations of $B_K$ using
lattice gauge theories, 1/N expansion and QCD sum rules of
various sorts, give consistently $B_K < 1$. The range
\begin{equation}
B_K = 0.7 \pm 0.2
\label{F4}
\end{equation}
is a good representaton of the 1/N \cite{BBG}, lattice results
\cite{GUP} and QCD sum rules results \cite{DEK}.
The QCD hadron duality approach gives
somewhat lower values $B_K = 0.4 \pm 0.1$ \cite{PIR}.
Finally the
leading term in the chiral perturbation theory in the strict
$SU(3)$ limit gives $B_K =
1/3$ \cite{DON}.

\item The next-to-leading QCD corrections to the QCD factor
$\eta_2$ (see below) \cite{BJW} decrease its value from
0.62 to 0.57 making the lower bound on $m_t$ somewhat
stronger.
\end{itemize}

In the coming years further improvements
on $\mid V_{cb}\mid, \mid V_{ub}/V_{cb} \mid$ and $B_K$ are
to be expected. It is then of interest to reanalyse the
bound (1) as a function of these three parameters. As a
byproduct of our analysis we will derive an approximate
analytic expression for the function $F$. Our analysis
complements the numerous analyses of the unitarity triangle
where the lower bound on $m_t$ has not been addressed.

Formulated in terms of the Wolfenstein parameters $\lambda,
A, \rho$ and $\eta$ the lower bound on $m_t$ arises from
the measured value of
$\epsilon_K$ roughly as follows. The usual box diagram
calculation together with the experimental value for
$\epsilon_K$ specifies the following hyperbola in the $(\rho,
\eta > 0)$ plane \cite{HAR,BHA}
\begin{equation}
\eta \left[(1-\rho) A^2 \eta_2 S(x_t)
+ \lambda^{-4} P_0\right] A^2 B_K = 0.223
\label{F5}
\end{equation}
Here
\begin{equation}
A \equiv \mid V_{cb}\mid/\lambda^2\qquad \lambda = \mid V_{us}\mid
= 0.22
\label{F6}
\end{equation}
\begin{equation}
P_0 = \eta_3 S(x_c,x_t) - \eta_1 x_c
\label{F7}
\end{equation}
\begin{equation}
S(x_c,x_t) = x_c \left[ \ln \frac{x_t}{x_c} - \frac{3x_t}{4(1-x_t)}
\left(1+ \frac{x_t}{1-x_t}\ln x_t\right)\right]
\label{F8}
\end{equation}
\begin{equation}
S(x_t) = x_t \left[ \frac{1}{4} + \frac{9}{4} \frac {1}{(1-x_t)} - \frac
{3}{2} \frac{1}{(1-x_t)^2}\right] + \frac{3}{2}
\left[ \frac{x_t}{x_t-1}\right]^3 \ln x_t
\label{F9}
\end{equation}
where $x_i = m^2_i/M^2_W$. $B_K$
is the renormalization group
invariant
non-perturbative parameter describing the size of
$<\bar{K^0} \mid (\bar s d)_{V-A}(\bar s d)_{V-A}\mid K^0>$
and $\eta_i$ represent QCD corrections to the box diagrams.

On the other hand the experimental value of $\mid
V_{ub}/V_{cb}\mid$ determines a circle of the radius
\begin{equation}
R_b=\frac{1}{\lambda} \left|\frac{V_{ub}}{V_{cb}}\right|
= \left[{\rho^2 + \eta^2}\right]^{1/2}
\label{F10}
\end{equation}
centered at $(\rho,\eta) = (0,0)$.
The hyperbola (5) intersects the circle
(10) in two points, one for $\rho < 0$ and the other
for $\rho > 0$. With decreasing $A$ (or $\mid V_{cb}\mid$), $B_K$ and
$m_t$, the hyperbola (5) moves away from the origin
of the $(\rho,\eta)$ plane. For sufficiently low
$A, B_K, m_t$ and $R_b$ the hyperbola and the circle only
touch each other. This way a lower bound for $m_t$ as a function
of $B_K, V_{cb}$ and $\mid V_{ub}/V_{cb}\mid$ can be found.

For the numerical evaluation of this bound it is
better to use the exact standard parametrization of the CKM
matrix \cite{PDG} and a more accurate formula
for $\epsilon_K$ given e.g.
in (5.22) of ref.\cite{BHA}. However, the formulation given above
allows to find a simple analytic expression
for $(m_t)_{min}$ which to an accuracy better than 2\%
reproduces the results of a more elaborate analysis.

In order to find a formula for $(m_t)_{min}$ which is
simple and accurate we proceed as follows. We note first
that in the ``exact'' numerical analysis which uses the
standard parametrization of the CKM matrix the lower bound
on $m_t$ corresponds to $sin \delta = 1$ i.e. to $\rho = 0$.
Equations (5) and (10) being approximations give
generally $(m_t)_{min}$ at a small negative value of $\rho$
rather than at $\rho = 0$. In spite of this the most
efficient strategy is to set $\rho = 0$ in (5) and
(10) to find the following condition for $(m_t)_{min}$:
\begin{equation}
S(x_t) = \frac{1}{A^2\eta_2} \left[\frac{0.223}{A^2B_KR_b} -
\frac{P_0}{\lambda^4}\right]
\label{F11}
\end{equation}
We use next \cite{BHA}
\begin{equation}
S(x_t) = 0.784~~ x^{0.76}_t
\label{F12}
\end{equation}
which is an excellent approximation of (9) in the range
$100 \le m_t \le 300~GeV$. In the same range of $m_t$,
$P_0$ takes the values $6.25~~ 10^{-4} \le P_0 \le 7.59~~ 10^{-4}$
where we
have used $\eta_1 = 0.85, \eta_3 = 0.36$ and $m_c = 1.4~GeV$. In
view of this weak $m_t$-dependence compared to (12), $P_0$ can
be approximated by a constant. This constant can be
chosen in such a way that the analytic bound reproduces the
results of a more accurate numerical analysis as good as
possible. $P_0 = 6.25~~ 10^{-4}$ turns out to be a good choice.

{}From (11) and (12) we finally obtain the analytic
lower bound on $m_t$:
\begin{equation}
(m_t)_{min} = M_W \left[\frac{1}{2A^2}\left(
\frac{1}{A^2B_KR_b} -1.2\right)\right]^{0.658}
\label{13}
\end{equation}
which is the main formula of this letter. To this end
we have set $\eta_2 = 0.57$ [16]. This QCD factor is so
defined that the resulting $m_t$ is the current top quark
mass normalized at $\mu = m_t$. The $m_t$ dependence
of $\eta_2$ can be neglected. Formula (13) together with
(6) and (10) gives $(m_t)_{min}$ as a function of
$\mid V_{cb}\mid, \mid V_{ub}/V_{cb}\mid$ and $B_K$.
It is evident from (13)
that this bound increases with decreasing $A$ (or $\mid
V_{cb}\mid ), B_K$ and $R_b$ (or $\mid V_{ub}/V_{cb}\mid )$.
We note a very strong dependence on A.
We have checked that in the ranges of $\mid V_{cb} \mid,
\mid V_{ub}/V_{cb} \mid$ and $B_K$ used below, our analytic
bound gives a very good representation (generally to better
than 2\% accuracy) of a more elaborate numerical analysis
which uses accurate formulae of ref.\cite{BHA}

In order to get better acquainted with the bound (13) we note
that for $0.035\leq
\mid~V_{cb} \mid \leq 0.043$ and
$0.06 \leq \mid V_{ub}/V_{cb} \mid \leq 0.10$ we have
$0.72 \leq A \leq 0.89$ and $0.27 \leq R_b \leq 0.45$ respectively.
Consequently for $B_K$ given in (4) we find $3.1 \leq 1/(A^2B_KR_b)
\leq 14.3$. For central values $B_K = 0.7, \mid V_{ub}/V_{cb}\mid
= 0.08$ and $\mid V_{cb}\mid = 0.038$ we have $1/(A^2B_KR_b) = 6.5$.

The bound in (13) is very interesting as it results
from
a different sector of the standard model than the
corresponding bound on $m_t$ obtained from high precision
electroweak
studies at LEP and Tevatron. Whereas the fate of the latter
bounds depends sensitively on the precise measurements of
$M_W, \Gamma_Z, sin^2 \Theta_W$, etc., the lower bound in
(13) is subject to our knowledge of $\mid V_{cb} \mid,
\mid V_{ub}/V_{cb}\mid $ and $B_K$.

In figs. 1-3 we plot $(m_t)_{min}$ as a function of $\mid
V_{cb}\mid$ for different values of $B_K$ and $\mid
V_{ub}/V_{cb} \mid$. We have set $M_W = 80.~GeV$.
These plots are self explanatory.
Therefore we only make a few comments.

We note that for $B_K \le 0.5$ the lowest value of $m_t$
consistent with the observed CP violation is generally
substantially larger than 200 GeV. For $B_K$ in the range
(4), the values of $(m_t)_{min}$ are compatible with the
restrictions on $m_t~~ (150 \pm 40~GeV)$ coming from
electroweak studies, although for $\mid V_{ub}/V_{cb}\mid \le
0.08,~
\mid V_{cb} \mid \le 0.038$ and $B_K \le 0.7$ again $(m_t)_{min} >
200~GeV$ is favoured. In such a situation a discovery of
the top quark with $m_t \approx 150~GeV$ would certainly
require new positive contributions to $\epsilon_K$ in order
to lower the bound on $m_t$ coming from the observed CP
violation. Charged Higgs exchanges and/or various
supersymmetry contributions to $\epsilon_K$ would be the
prime candidates but such an analysis is beyond the scope of
this letter.

It is interesting to compare these bounds with those which
can be obtained from $B^o-\bar B^o$ mixing alone. The
experimental knowledge of the $B^o-\bar B^o$ mixing described
by the parameter $x_d = \Delta M/\Gamma_B$ determines in the
$(\rho,\eta)$ plane a circle centered at $(\rho,\eta) =
(1,0)$ and having the radius
\begin{equation}
R_t = \frac{1}{\lambda}\left| \frac{V_{td}}{V_{cb}}\right|
= \left[ (1-\rho)^2 + \eta^2\right]^{1/2}
\label{F14}
\end{equation}
Using the usual formulae for box diagrams with top quark exchanges
\cite{BHA} it is straightforward to find the value of $m_t$ consistent
with $x_d$ as a function of $R_t$, the $B$-meson decay constant
$F_B$, the parameter $B_B$ analogous to $B_K$ and
$\kappa$ of (2).
Defining
\begin{equation}
r \equiv\left[ {\sqrt\frac{x_d}{0.67}} \left[\frac{200 MeV}{F_B
\sqrt {B_B}}\right]\left[\frac{0.038}{\kappa}\right]\sqrt
\frac{0.55}{\eta_{B}}\right]^{1.32}
\label{F15}
\end{equation}
we find
\begin{equation}
m_t = r~R^{-1.32}_t 179.4~GeV
\label{R16}
\end{equation}
Here $\eta_B$ is the QCD factor analogous to $\eta_2$ and calculated
to be $\eta_B = 0.55$ \cite{BJW}.
Since (10) and (14) must be consistent with each
other the lower bound on $m_t$ from $B^o-\bar B^o$ mixing without any
constraint from $\epsilon_K$ is found by setting $\eta = 0$ and
$\rho = - R_b$ which implies $R_t = 1 + R_b$. This gives then
\begin{equation}
(m_t)_{min} = \left\{ \begin{array}{r}
r \cdot 110~ GeV \quad \mid V_{ub}/V_{cb}\mid = 0.10\\ \nonumber
r \cdot 120~ GeV \quad \mid V_{ub}/V_{cb}\mid = 0.08\\ \label{F17}
r \cdot 131~ GeV \quad \mid V_{ub}/V_{cb}\mid = 0.06 \nonumber
\end{array}\right. \end{equation}
We observe
that this bound depends much weaker on $\mid
V_{ub}/V_{cb}\mid$ than the bound on
$\epsilon_K$ given in (13).

The fate of the lower bound from $B^o-\bar B^o$ mixing
depends on the ratio $r$ in (15). With $x_d =0.67 \pm
0.10$ \cite{DAN},
$\kappa = 0.038 \pm 0.006$ and $F_B{\sqrt B_B} = 200 \pm 40
MeV$ \cite{CTS} one has the range $0.58 \le r \le 1.85$ i.e. a large
uncertainty. For central values of the parameters the lower
bound from $B^o-\bar B^o$ mixing is generally weaker than
from $\epsilon_K$. It should however be emphasized that the
lower bound from $B^o-\bar B^0$ mixing in 1993 is
substantially higher than the bounds found in the 80's \cite{AF}.
This is primarily due to the values of $\kappa$ and $\eta_B$
which were substantially higher in those days. For instance
with $V_{cb} \simeq 0.05$ and $\tau_B \simeq 1.2 ps$ one had
$\kappa \simeq 0.046$. Taking in addition $\eta_B\simeq 0.85$
used in older analyses decreases the lower bound on $m_t$ by
a factor of 1.7. This factor is somewhat reduced by the
higher values of $F_B$ found at present.

In this letter we have emphasized that the lower
bounds on $m_t$ coming from $\epsilon_K$ and
$B^o-\bar B^o$ mixing increased considerably during the
last years. In particular we have found that for central
values of the parameters the lower bound on $m_t$ from
$\epsilon_K$ appears to be higher than the estimates of $m_t$
from the precision electroweak studies at LEP. Are these some
hints for the physics beyond the standard model?
In order to answer this question continuous efforts should
be made to decrease the uncertainties in $\mid V_{cb}\mid,
\mid V_{ub}/V_{cb}\mid, B_K$ and $F_B$. It will be exciting
to watch in the coming years the developments in $B$-decays,
in electroweak studies and in particular in top quark
searches. We hope that figs. 1-3 and the analytic lower
bounds on $m_t$ derived here will help to see quickly where
we stand.

\bigskip
I would like to thank Gerhard Buchalla and Gaby Ostermaier
for a careful reading of the manuscript. I also thank Dean Karlen and
Sheldon Stone
for e-mail messages.

\large{\bf Figure Captions}
\normalsize
\bigskip
\begin{itemize}
\item[Fig.1] The lower bound on $m_t$ from $\epsilon_K$ as a
function of $\mid V_{cb}\mid$ for different $B_K$ and
$\mid V_{ub}/V_{cb}\mid = 0.06.$
\item[Fig.2] Same as fig.1 but for $\mid V_{ub}/V_{cb}\mid
=0.08.$
\item[Fig.3] Same as fig.1 but for $\mid V_{ub}/V_{cb}\mid
=0.10.$
\end{itemize}


\begin{thebibliography}{999}
\bibitem{PHG} P.H. Ginsparg, S.L. Glashow and M.B. Wise,
{\sl Phys.Rev.Lett}, {\bf 50}, (1983) 1415.
\bibitem{AJB} A.J. Buras, W. S{\l}ominski and H. Steger,
{\sl Nucl.Phys.} {\bf B238} (1984) 529.
\bibitem{SLS} S.L. Stone, in B-Decays, edited by S.L. Stone
(World Scientific, Singapore, 1992) p. 210.
\bibitem{NA} D. Karlen, ``B-Hadron Lifetimes'', talk presented
at the Heavy Flavours Conference, Montreal, July 1993.
\bibitem{NEU} M. Neubert, {\sl Phys.Lett.} {\bf 264B} (1991) 455;
{\sl Phys.Rev.} {\bf D46} (1992) 2212.
\bibitem{MN} M. Neubert, ``Heavy Quark Symmetry'', SLAC-PUB 6263
(1993).
\bibitem{BB} P. Ball, {\sl Phys.Lett.} {\bf 281B} (1992) 133.
\bibitem{ST} G. Burdman, {\sl Phys.Lett.} {\bf 284B} (1992) 133.
\bibitem{BBD} P. Ball, V.M. Braun, H.G. Dosch, TUM-T31-31/92
to appear in Phys.Rev.D.
\bibitem{ART} M. Artuso, Syracuse Preprint, HEPSY-1-93.
\bibitem{BBG} W.A. Bardeen, A.J. Buras and J.-M. G\'erard,
{\sl Phys.Lett.} {\bf 211B} (1988) 343.
\bibitem{GUP} R. Gupta, G.W.Kilcup and R.S. Sharpe,
{\sl Nucl.Phys.} B (Proc.Suppl.) {\bf 26} (1992) 197;
M.B. Gavela et al. {\sl Nucl.Phys.}{\bf B306} (1988) 138;
C. Bernard and A. Soni, {\sl Nucl.Phys.} B (Proc.Suppl.) {\bf 17}
(1990) 495; N. Ishizuka et al., {\sl Phys.Rev.Lett.}{\bf 71}
(1993) 24.
\bibitem{DEK} R. Decker, in Proceedings of the Ringberg Workshop on
``Hadronic Matrix Elements and Weak Decays'', {\sl Nucl.Phys.} B
(Proc.Suppl.) {\bf 7A} (1989).
\bibitem{PIR} A. Pich and E. de Rafael, {\sl Nucl.Phys.} {\bf B358}
(1991) 311.
\bibitem{DON} J.F. Donoghue, E. Golowich and B.R. Holstein,
{\sl Phys.Lett.} {\bf 119B} (1982) 412.
\bibitem{BJW} A.J. Buras, M. Jamin and P.H. Weisz,
{\sl Nucl.Phys.} {\bf 347} (1990) 491.
\bibitem{HAR} G.R. Harris and J.L. Rosner, {\sl Phys.Rev.}
{\bf D45} (1992) 946.
\bibitem{BHA} A.J. Buras and M.K. Harlander, A Top Quark Story,
in Heavy Flavors, eds. A.J. Buras and M. Lindner, World
Scientific, Singapore (1992) p. 58.
\bibitem{PDG} Particle Data Group, {\sl Phys.Rev.} {\bf D45}
(1992), No 11, part II.
\bibitem{DAN} D. Danilov, in proceedings of the International
Lepton-Photon Symposium (Geneva, 1991).
\bibitem{CTS} C.T. Sachrajda in Heavy Flavours, edited by
A.J. Buras and M. Lindner (World Scientific, Singapore, 1992),
p.415;\\
E. Bagan et al. {\sl Phys.Lett.} {\bf B278} (1992) 457;\\
C. Bernard et al. UW/PT-93-0.6;\\
M. Neubert, {\sl Phys.Rev.} {\bf D45} (1992) 2451.
\bibitem{AF} G. Altarelli and P.J. Franzini, {\sl Z.Phys.} {\bf C37}
(1988) 271.
\end{thebibliography}
\end{document}